\documentclass[preprint2]{aastex}

\newcommand{\xmm}{{\em XMM-Newton}}
\newcommand{\wo}{WR\,142}

\newcommand{\Lbol}{\mbox{$L_{\rm bol}$}}

\newcommand{\Lx}{\mbox{$L_{\rm X}$}}

\newcommand{\vinf}{\mbox{$v_\infty$}}
\newcommand{\mdot}{\mbox{$\dot{M}$}}   

\newcommand{\lsim}{\raisebox{-.4ex}{$\stackrel{<}{\scriptstyle \sim}$}}

\def \etal   {\hbox{et~al.\/}}
\def\changed{}
\def\nchange{}
\shorttitle{X-ray emission from a WO type star}
\shortauthors{Oskinova et al.}

\begin{document}
\title{Discovery of X-ray Emission from the Wolf-Rayet star WR~142 of oxygen 
subtype}


%\author{L. M. Oskinova\altaffilmark{1}, W.-R. Hamann\altaffilmark{1}}
% No afiliation exponents anymore??

\author{L.~M.~Oskinova, W.-R.~Hamann, A.~Feldmeier}
\affil{Institute for Physics and Astronomy, University Potsdam, 
14476 Potsdam, Germany\\
\email{lida@astro.physik.uni-potsdam.de}}
\author{R.~Ignace}
\affil{Department of Physics and Astronomy, East Tennessee State University,
Johnson City, TN 37614, USA}
\author{Y.-H.~Chu}
\affil{Department of Astronomy, University of Illinois, 
1002 West Green Street, Urbana, IL 61801, USA}


%\altaffiltext{1}{Visiting Astronomer, Cerro Tololo Inter-American Observatory.
%CTIO is operated by AURA, Inc.\ under contract to the National Science
%\altaffiltext{2}{Society of Fellows, Harvard University.}
%\altaffiltext{3}{present address: Center for Astrophysics,
%    60 Garden Street, Cambridge, MA 02138}
%\altaffiltext{4}{Visiting Programmer, Space Telescope Science Institute}
%\altaffiltext{5}{Patron, Alonso's Bar and Grill}

\begin{abstract}  We report the discovery of weak yet hard X-ray
emission from the Wolf-Rayet (WR) star \wo\ with the \xmm\ X-ray
telescope. Being of spectral subtype WO2, \wo\ is a massive star in a
very advanced evolutionary stage, short before its explosion as a
supernova or $\gamma$-ray burst. This is the first detection of X-ray
emission from a WO-type star.  {\changed We rule out any serendipitous
X-ray sources within $\approx 1''$ of \wo.}
{\nchange  \wo\ has an X-ray luminosity of $L_{\rm
X}=7\times10^{30}$\,erg\,s$^{-1}$, which constitutes only $\lsim
10^{-8}$ of its bolometric luminosity.}  The hard X-ray spectrum
suggests a plasma temperature of about 100\,MK.  Commonly, X-ray
emission from stellar winds is attributed to embedded shocks due to
the intrinsic instability of the radiation driving. From qualitative
considerations we conclude that this mechanism cannot account for the
hardness of the observed radiation. There are no hints for a binary
companion. Therefore the only remaining, albeit speculative
explanation must refer to magnetic activity. Possibly related, \wo\
seems to rotate extremely fast, as indicated by the unusually round
profiles of its optical emission lines. Our detection implies that the
wind of \wo\ must be relatively transparent to X-rays, which can be
due to strong wind ionization, wind clumping, or non-spherical
geometry from rapid rotation.  \end{abstract}



\section{Introduction}

Stars of the Wolf-Rayet (WR) type have a highly peculiar chemical
composition and very strong stellar winds. The WR spectra are sorted
into three subclasses: WN, WC, and WO, according to the dominance of
nitrogen, carbon, and oxygen emission lines, respectively. These spectral
types correspond to evolutionary stages \citep{con83}.  The largest C+O
abundance and the fastest stellar winds are observed in WO type stars,
that represent the final evolutionary stage of a massive star prior to
its explosion as a type~Ic supernova or $\gamma$-ray burst \citep{hir05}.

The WR stars continue to challenge our understanding of line-driven
winds.  \citet{sch96} pointed out the importance of the iron opacity
for the acceleration of WR winds.  The first hydrodynamical model for
a WR wind was presented by \citet{gr05}. In their simulation the mass
loss is initiated at high optical depth by the so-called ``iron bump'' 
in the opacity. It was thus demonstrated that WR-type
winds can be driven by radiation pressure.

It has long been known that line-driven winds are subject to an
instability that can lead to strong shocks
\citep{lw80}.  These shocks are thought to explain the X-ray emissions
from O~star winds, as predicted by time-dependent hydrodynamic
modeling \citep{ow88,feld97} and largely confirmed by observations
\citep{kr03,os06,zh07,wc07}. The growth of instability in WR winds was 
investigated by \citet{go95}. They found that despite of 
damping effects due to the multi-line scattering, the instability
remains effective. Therefore, X-ray emission from wind shocks could,
in principle, be expected in WR~winds, a conjecture that has not been
yet tested by time-dependent hydrodynamic simulations.

Significant observational effort has been made to study the X-ray
emission of WR stars. {\changed \citet{wl86}, \citet{pol87},
\citet{pol95}, \citet{os05} presented X-ray observations of
Galactic WR stars.} A survey of X-ray emission from WR stars in the
Magellanic Clouds was conducted by \citet{gu08a,gu08b}. \citet{ig00}
and \citet{os05} demonstrated that X-ray properties of single WR stars
differ from those of O~stars.  {\changed Whereas O~stars display a
trend in which the ratio of the X-ray to the bolometric luminosity
$\Lx/L_{\rm bol}$ has a typical value of $10^{-7}$
\citep{longw80,berg97,sana06}}, this trend is not observed 
in the case of WR~stars.

Observations with the \xmm\ and {\em Chandra} X-ray telescopes  
established that some bona fide single WN stars are X-ray active
\citep{sk1,sk2,ig03, os05}, while others are apparently
not \citep{os05,gos05}. \citet{os03} found that no single WC~star had
been conclusively detected at X-ray energies, a result that continues
to hold.  Among the WR subclasses, only the WO-type stars have not
been observed in X-rays so far. In this {\em Letter} we present the
\xmm\ observations of the closest WO type star \wo.  The star is
introduced in Sect.\,2, and its \xmm\ observations are described in
Sect.\,3. The implications of the \xmm\ detection of \wo\ are
discussed in Sect.\,4.

\section{The WO-type Star \wo}
\label{sect:thestar}

\wo, also named Sand\,5 and St\,3, has a spectrum characteristic for
the spectral subtype WO2 \citep{bar82,kin94}.  The optical spectrum of
\wo\ was discussed by \citet{p97}, who also noticed some
line variability.

\begin{figure}
\epsscale{1.7}
\plotone{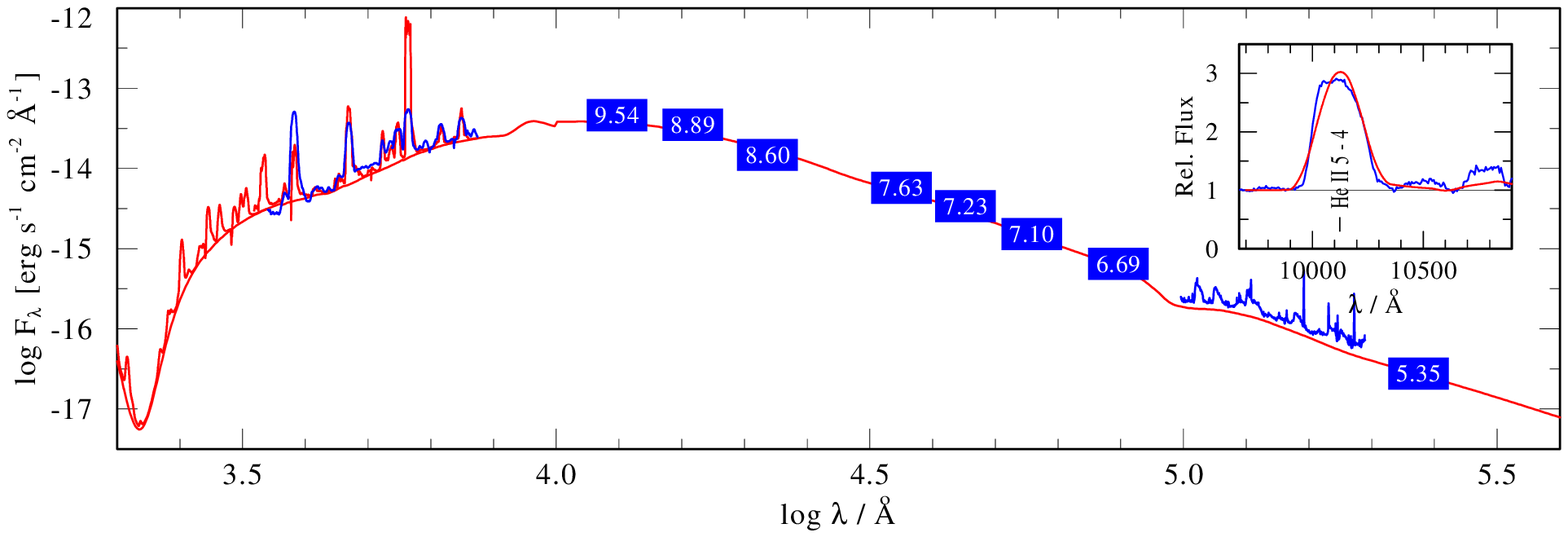}
\caption{Spectral energy distribution of WR\,142. {\changed
Boxes give observed 2MASS, {\em Spitzer} IRAC and MIPS magnitudes 
(labels). Observed optic and IR spectra are shown by thin blue lines.} 
The red line is the continuum flux (plus lines in the UV and
optical) from a PoWR model with $\log L/L_\odot = 5.35$, $T_\ast =
160$kK, $\mdot=10^{-5.1} M_\odot {\rm yr}^{-1}$ reddened with
$E_{B-V}=1.73$ mag. The model with $\log L/L_\odot = 5.65$ and
$\mdot=10^{-5.4} M_\odot {\rm yr}^{-1}$ gives an equally good fit. 
A typical He\,{\sc ii} line is shown in the inset
to illustrate the effect of extreme rotational broadening -- the observed line 
(blue) is compared to the synthetic line convolved with 
$v\sin{\rm i}=4000$\,km\,s$^{-1}$ (red).    
\label{fig:sed}}
\end{figure}

Figure\,\ref{fig:sed} shows the spectral energy distribution (SED) of
\wo\ together with model calculated with the Potsdam Wolf-Rayet
(PoWR) model atmosphere code \citep{gkh02}. {\changed Photometric IR
measurements } are plotted together with our optical spectrum and the
{\em Spitzer} IRS mid-IR spectrum.  We have adopted parameters typical
for a WO star: stellar temperature $T_\ast=150$\,kK, ``transformed
radius''
\citep[cf.][]{gkh02} $R_{\rm t} = 2R_\odot$, and a composition of 
40\% carbon, 30\% oxygen and 30\% helium (by mass). WR\,142 is assumed
to be a member of the open cluster Berkeley\,87 at a distance of
$d=1.23$\,kpc \citep{tur06}.  Based on this preliminary model, 
the fit of the photometric
observations requires an interstellar reddening of $E_{B-V} = 1.7$mag
and a stellar luminosity of $\log \Lbol/L_\odot = 5.35$, implying a
stellar radius of only $R_\ast = 0.6\,R_\odot$.  The corresponding 
mass-loss rate is about $10^{-5.1} M_\odot {\rm
yr}^{-1}$ for an adopted microclumping volume filling factor of 0.1.

{\changed The adopted model} does not provide an entirely satisfactory
fit to the line spectra. For example, the model does not match the
huge O\,{\sc vi} emission at 3811,\,3834\,\AA, a problem also
experienced by \citet{cr00} while reproducing these lines for Sand\,2
with the {\sc cmfgen} code. We also fitted the SED shown in
Fig.\,\ref{fig:sed} with a model that has a mass-loss rate lower by a
factor of two, $10^{-5.4} M_\odot {\rm yr}^{-1}$, and higher
bolometric luminosity, $\log \Lbol/L_\odot = 5.65$. This model fits
the SED in Fig.\,\ref{fig:sed} equally well. Figure\,\ref{fig:rtau1}
shows the radius of unity optical depth plotted as function of
wavelength in the X-ray range for both models. The stronger wind is
opaque even to hard X-rays, but the thinner wind is largely
transparent, because its higher ionization reduces the X-ray absorbing
ions.  The same could happen in denser, but hotter models.  Thus our
ability to predict the influence of wind photo-absorption is somewhat
limited, owing to ambiguities in the ionization state of metals and
uncertainty in the mass-loss rate.

The profiles of the emission lines in the spectrum of WR\,142 are very
broad. Assuming that the line widths correspond to the wind terminal
velocity, the velocity of $\vinf \approx 5500$\,km\,s$^{-1}$ would be
deduced.  However, the profile shapes of almost all lines are much more
round than the roughly Gaussian shapes usually  seen in WR spectra. Such
round profiles cannot be reproduced by the standard models. It is
tempting to reproduce these profiles by convolution with the
semi-ellipse for rotational broadening, albeit  rotating stellar winds
certainly require a more sophisticated treatment which has not been
accomplished yet.  If rotation is the cause for the round profiles,  the
projected rotation speed must be comparable to \vinf, i.e.\
the star would rotate near to break-up with $v_{\rm rot}\sin{i}
\lsim 4000$\,km\,s$^{-1}$ (see inset in Fig.\,\ref{fig:sed}).
Interestingly, a similar suggestion has been made for the hottest and
the most compact Galactic WN star, WR\,2, ($T_\ast = 140\,{\rm kK},
R_\ast = 0.89\,R_\odot, v_{\rm rot}\sin{i}\approx 2000 {\rm
km\,s}^{-1}$) \citep{hgl06}. 

%The importance of high rotational
%velocities for the evolution of WN stars has been highlighted by
%\citet{mar08}.

\section{Observations}
\label{sect:obs}

\begin{figure}
\epsscale{.7}
\plotone{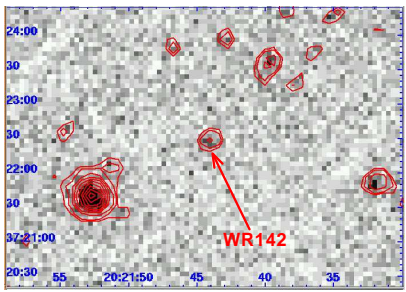}
\caption{Part of the merged \xmm\ EPIC image (0.2-12.0\,keV) with 
over-plotted contours. Image size is  $5.7' \times 4.1'$.
\wo\ is marked by an arrow.
The coordinates are equatorial (J2000). North up,  east left. 
\label{fig:epic}}
\end{figure}

\wo\ was observed by \xmm\ during two consecutive satellite orbits
(ObsId\,0550220101 and ObsId\,0550220201).  The data were merged and
analyzed using the latest versions of software {\sc
sas}\,8.0.0. {\changed After the high background level time intervals
have been rejected,} the combined exposure time of all detectors was
$\approx 100$\,ks; an EPIC image of \wo\ and its surroundings is shown
in Fig.\,\ref{fig:epic}.
{\changed 
\wo\ is detected  with a 5$\sigma$ confidence level in all \xmm\ EPIC 
detectors using standard source-detection algorithms. }  We define a
``hardness ratio HR'' as ${\rm HR}= ({\rm N}_{\rm hard}-{\rm N}_{\rm
soft}) ({\rm N}_{\rm hard}+{\rm N}_{\rm soft})^{-1}$, where ${\rm
N}_{\rm soft}$ is the number of counts in the 0.25-2.0\,keV band and
${\rm N}_{\rm hard}$ in the 2.0-12.0\,keV band.  For \wo\ we find
HR$\approx 0.57$.  For comparison, the hardness ratio of the WN star
WR\,1, which has a reddening similar to \wo, is HR(WR\,1)$\approx
-0.9$, while more reddened WR\,110 has HR(WR\,110)$\approx -0.4$ and
less reddened WR\,6 has HR(WR\,6)$\approx -0.6$. Thus X-ray emission
from \wo\ is the hardest among putatively single WR stars.  The count
rates in the 0.25\,keV-12.0\,keV band are $(1.89\pm 0.34)\times
10^{-3}$\,c\,s$^{-1}$ for EPIC MOS1+2 and $(3.80
\pm 0.84)\times 10^{-3}$\,c\,s$^{-1}$ for EPIC PN cameras.  
{\changed Assuming a two-temperature thermal plasma model 
($kT_1=0.3\,{\rm keV}, kT_2=10$\,keV)}, 
{\nchange the observed X-ray flux of \wo\ is 
$F_{\rm X}=4\pm 2 \times 10^{-14}$\,erg\,s$^{-1}$\,cm$^{-2}$.} The reddening 
towards \wo\ is known from the analysis of its optical spectrum, and the
distance is known from its cluster membership. {\nchange The X-ray 
luminosity of \wo\ is thus $L_{\rm X}\approx 7\times 10^{30}$\,erg\,s$^{-1}$, 
or $\log{L_{\rm X}/L_{\rm bol}}\approx -8$.}

The angular resolution of \xmm\ is $\lsim\ 6''$. To exclude the
potential confusion with a source in close vicinity of \wo, we
inspected optical and infra-red images with higher angular
resolution. According to the USNO-B1.0 catalog \citep{usno}, the
closest object to \wo\ is located $8''$ away.  The optical monitor
(OM) on board of \xmm\ provides images with an angular resolution of
$\approx 1''$.  The OM images in four filters did not detect 
any objects closer than $8''$ around
\wo.  The {\em Spitzer} space telescope has observed the Berkeley\,87
cluster in the infra-red (IR). With an angular resolution of $\lsim\
1''$, the {\em Spitzer} IRAC camera took images in four channels.  No
point sources within $8''$ from \wo\ were detected in any of the IRAC
channels. It is extremely unlikely that any X-ray source is not detectable
in either the optical or IR. We therefore rule out any
serendipitous X-ray sources within $\approx 1''$ of \wo\ that might get
confused as our target.

In addition, there is no evidence to suggest that the observed X-rays
originate from a wind blown bubble around \wo. Although theoretically
expected, there is a dearth of WR stars with detected diffuse X-ray
emission from their wind-blown bubbles \citep{chu03,wr05}. The only
two detected hot bubbles show a limb-brightened morphology and are
extended on the scale of parsecs. On the contrary, we confidently
detect a point source at the position of \wo.

\begin{figure}
\epsscale{0.7}
\plotone{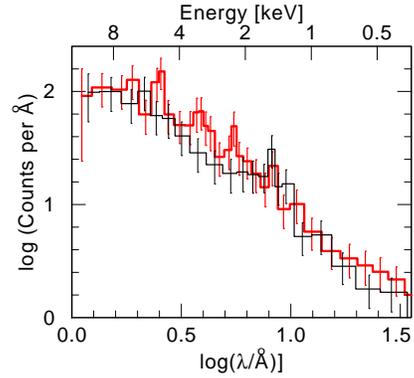}
\caption{Thick red line: \xmm\ MOS-2 spectrum of \wo\ 
where background has {\em not} been subtracted. Thin blackline: MOS-2 spectrum 
of a background region. The error bars give $2\sigma$. \label{fig:sp}}
\end{figure}

\begin{figure}
%\plotone{mos2-spectrum.ps}
\plotone{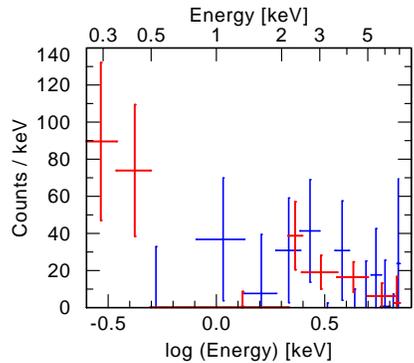}
\caption{\xmm\ MOS-2 (red) and PN (blue) spectra of \wo\  with 
subtracted background. The error bars give 2$\sigma$ } 
\label{fig:m2sp}
\end{figure}

The spectra of \wo\ were extracted from the $15''$ region. The small
number of counts obtained in our observation does not allow for
quantitative spectral analysis, {\changed but interesting conclusions
can be made about the gross energy distribution.}
Figure\,\ref{fig:sp} shows the X-ray spectrum of \wo\ before a
background subtraction, together with the spectrum of a nearby
background region normalized to the same area. The emission from \wo\
dominates over the background at $2-7$\,\AA\ ($1.7-6$\,keV), while
longwards it dives below the background until about $\approx 20$\,\AA\
(0.6\,keV) where it rises above the background again.  {\changed The
presence of ``a dip'' between $\approx 7-12$\,\AA\ ($0.8-1$\,keV) is
also seen in the background-subtracted PN and MOS-2 spectra shown in
Fig.\,\ref{fig:m2sp}. The appearance of the dip may be due to
unresolved strong emission lines. However,} the wavelength of the dip
coincides well with the K-shell edges of oxygen
\citep{ver95}, where our stellar wind models predict an \ absorption
{\em maximum} (see Fig.\,\ref{fig:rtau1}). Therefore it is tempting to
attribute the observed dip to the oxygen K-shell absorption -- to
confirm this identification, data of higher quality will be
required. Assuming the presence of K-shell of oxygen, the thermal
plasma model fit to our low S/N data indicates presence of plasma with
temperatures spanning from 1\,MK to 100\,MK.

\section{On The Origin of X-ray Emission from \wo }

The confident detection of weak hard X-rays from \wo\ prompts us to
re-consider the previous non-detections of X-ray emission from WR stars.
{\nchange The X-ray luminosity of \wo\ of $L_{\rm X}\approx 7\times
10^{30}$\,erg\,s$^{-1}$ is comparable to present upper limits for the
non-detections \citep{os03,os05,gos05,sk06}.  Perhaps some WR stars that
are yet undetected in X-rays  may be weak sources  similar to \wo.} 
This raises the question, what is the mechanism responsible for the
generation of X-rays in WR~stars? Usual suspects are wind shocks and
magnetic fields.

The strong-shock condition predicts the peak temperature in the
shocked gas, $T_{\rm X}$, as
\begin{equation} 
kT_{\rm X}^{\rm max}=\frac{3}{16} \mu m_{\rm H} U^2,  
\label{eq:rh}
\end{equation} 
where $U$ is the velocity with which the gas rams into the shock (i.e.,
pre-shock velocity relative to the shock), and $\mu$ is the 
mean molecular weight per particle in the {\em post-shock} gas.  
As the  shock temperatures are very high, abundant species will be nearly
entirely ionized.  Therefore, $\mu$ depends only weakly on chemical
compositions. For example, $\mu \approx \frac{1}{2}$ for ionized H, 
while for fully ionized carbon, $\mu \approx \frac{2}{3}$. Thus, 
differences  in wind molecular weight between O and WR stars cannot 
strongly affect the temperatures of the shocked plasma in stars of 
these types.

{\nchange Note that Eq.\,(\ref{eq:rh}) gives an upper limit to the
temperature at the shock. Radiative cooling will restrict the hottest
temperatures to a limited part of the total emission measure. Moreover,
some fraction of the energy will be consumed by ionization processes. 
To roughly estimate the energy required to ionize WO wind material in
the post-shock zone we consider oxygen and assume that the leading ion
in the pre-shock material is O\,{\sc vi}, while in the post-shock plasma
it is O\,{\sc ix}.  A WO wind may contain 30\%\ of oxygen. The
ionization potentials are IP(O\,{\sc vi})$\approx 0.14$\,keV, IP(O\,{\sc
vii})$\approx 0.74$\,keV, and IP(O\,{\sc viii})$\approx 0.87$\,keV
\citep{cox}. The specific energy for ionization is $\epsilon=\Sigma
N_{\rm i}{\rm IP}_{\rm i}$, where $N_{\rm i}=X_{\rm i}(A_{\rm i}m_{\rm
H})^{-1}$ with $X_{\rm i}$ being the mass fraction of the element,
$A_{\rm i}$ its atomic weight, and $m_{\rm H}$ the atomic mass unit.
Inserting these numbers, the full ionization of oxygen requires
$\epsilon\approx 3\times 10^{13}$\,erg\,g$^{-1}$. This is small compared
to the kinetic energy of the wind ($5\times 10^{15}$\,erg\,g$^{-1}$ for
typical 1000\,km\,s$^{-1}$).  Thus the ionization is not a significant
cooling process even in metal-rich winds.}

In the time-dependent hydrodynamic simulations by \citet{feld97} the
velocity jump $U$ depends on the ratio between the period of the 
perturbations at the wind base, $T_{\rm c}$, and the flow time, $T_{\rm
flow}=R_\ast/v_\infty$. The former was estimated by the acoustic cut-off
period, $T_{\rm c}=a H^{-1}$, where $a$ is thermal speed and $H$ is the
pressure scale height. Assuming that the perturbation are seeded in the
hydrostatic layers within the star, $T_{\rm c}/T_{\rm flow} \propto
a\,v_{\rm esc}^{-1}$.  The thermal speed is lower in  hydrogen-deficient
WR atmospheres, while $v_{\rm esc}$ can be larger than in O stars, as it
is the case  for our program star \wo. Therefore, the velocity jumps in
WR wind shocks are  expected to be rather {\em smaller} than larger,
compared to O-star winds.

These qualitative considerations indicate that if the same shock
mechanism were in operation in WR and O star winds, the X-ray spectra
from WR stars would be softer than those of O stars. 
However, observations show the contrary -- the handful of single
WR stars with available spectral measurements all show X-ray emission
harder than typically found in O stars \citep{sk1,sk2,ig03}. This trend
is confirmed by our new data on \wo.

Relatively hard X-ray emission can occur in a binary system; however,
optical spectra of \wo\ show no contamination from an early-type companion.
A low-mass coronal companion to this initially very massive star is 
extremely unlikely \citep{l06}. 

{\changed \citet{bab97} proposed that if a stellar wind is
magnetically confined it can be strongly heated.  The magnetic field
locally dominates the bulk motions if the magnetic energy density
exceeds the wind kinetic energy density, $B^2/\mu_0>\rho v^2$. {\nchange  
At characteristic distance of 1\,$R_*$ from the photosphere, the velocity
is $v\approx 0.5\vinf\approx 2000$\,km\,s$^{-1}$, density
$\rho=\dot{M}/4\pi v(r)^2 r \approx 1 \times 10^{-10}$\,g\,cm$^{-3}$.
Therefore, $B(r=2R_*)>7$\,kG is required to control the \wo\
wind. For a dipole field, this implies that the magnetic field
strength at the surface is larger than 50\,kG.}  A field of such
strength is not unrealistic, since the radius of \wo\ is $\sim 50$
times smaller than that of OB supergiants, where surface fields of
100\,G have been observed \citep{br08}. One may speculate that even
less strong magnetic fields may lead to hard X-rays due to some type
of reconnection and heating processes as invoked to explain the high
X-ray temperatures observed in O~type stars \citep{wc07}.}

It is difficult to test {\changed where in the wind the observed}
X-rays have been produced. Since no high-resolution X-ray spectrum of
a single WR star exists, the {\em f-i-r} emission-line diagnostic that
has been used for O-star winds has not been applied to WR~stars.
X-rays produced close to the stellar core must travel through the bulk
of photo-absorbing wind. {\changed Tentative evidences for absorption
edges are found in the X-ray spectra of WR\,1
\citep{ig03} and \wo.}

\begin{figure}
\epsscale{.7}
\plotone{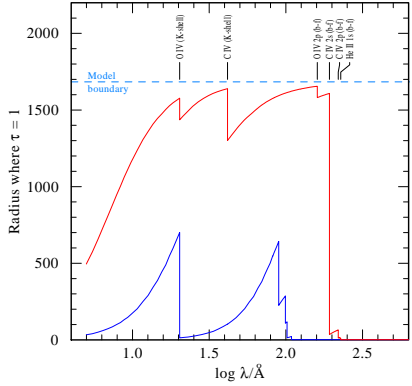}
\caption{Radius (in units of $R_\ast$) where the radial
optical depth becomes unity, as a function of wavelength in the X-ray
range. The red curve is for the same model for which the spectral energy
is shown in Fig.\,\ref{fig:sed}, while the blue curve is
for a model with half the mass-loss rate and otherwise similar
parameters. 
\label{fig:rtau1}}
\end{figure}

In the wind of \wo, the radius at which the photo-absorption to X-rays
is equal to unity is plotted in Fig.\,\ref{fig:rtau1} against photon
energy. The two shown models reproduce the stellar SED equally well.
One of the models is thick to X-rays for 1000's of $R_\ast$, and the
observed X-rays would have to emerge from the far outer regions of the
wind.  Origin of X-rays far out in the wind is favored in the
alternative scenario of X-ray emission from O stars \citep{pol07}.
{\changed However, a reduction in the mass-loss (a poorly constrained
parameter) by a factor of only two and/or a higher effective stellar
temperature result in a higher degree of wind ionization.  In this
case a fraction of X-rays formed deep in the wind could freely escape.
Wind clumping can further reduce wind attenuation \citep{foh03}.}

There is suggestive evidence that \wo\ is a fast rotator, as is the
case for the WN2 star WR\,2 (see Sect.\,\ref{sect:thestar}). The
detection of strong and hard ($\log{L_{\rm X}\approx
10^{32}}$\,erg\,s$^{-1}$, $T_{\rm X} \gg 10$\,MK) X-rays in WR\,2
\citep{sk08} likely indicates that a similar mechanism operates in these
hot, compact, fast rotating stars.  The lower X-ray luminosity of \wo\
compared to WR\,2 can reflect the higher absorption of X-rays in its
denser wind. Fast rotation can lead to a slow equatorial wind with
enhanced density, possibly forming an out-flowing disk, and a faster,
thinner polar wind \citep{ig96}.  If the round shape of the line
profiles found in both, \wo\ and WR\,2, is due to rotational
broadening, the inclination angle must be large.

Summarizing, we have reported the detection of weak, but hard X-rays
from \wo\ and speculate that their origin is connected with magnetic
activity of this far-evolved, compact WR  star. The specific shape of
its optical line profiles signals that \wo\ may be rotating at nearly
break-up velocity. In this case the spherical symmetry will be broken,
potentially affecting the X-ray production and absorption.  Observations
of better quality and progress in modeling are  certainly needed to
understand fully the high-energy processes  in the winds of massive
stars at late stages of their evolution.

\acknowledgments
Based on observations obtained with \xmm, an ESA science mission with
instruments and contributions directly funded by ESA Member States and
NASA.  This research has made use of NASA's Astrophysics Data System
Service and the SIMBAD database, operated at CDS, Strasbourg, France.
Special thanks to Robert Gruendl for providing the IR measurements.
{\changed The authors thank the referee M.~De Becker for important
comments that improved the paper.}  Funding for this research has been
provided by NASA grant NNX08AW84G (Y-HC and RI) and DLR grant
50\,OR\,0804 (LMO).

%------------------------------------------------------------------------

\clearpage

%---------------------------------------------------------------------
%---------------------------------------------------------------------

\clearpage

%---------------------------------------------------------------------
%---------------------------------------------------------------------

%---------------------------------------------------------------------
%---------------------------------------------------------------------

%\clearpage

\end{document}